%% file: KMB.tex
\documentclass[sigconf]{acmart}
\AtBeginDocument{%
  }

\usepackage{amsfonts}
\usepackage{textcomp}
\usepackage{url}
\usepackage{algpseudocode}
\usepackage{algorithm}
\usepackage{color}
\usepackage{xspace}
\usepackage{tikz}
\input{macros}

\newcommand{\algo}[1]{\textsc{#1}}
\newcommand{\new}[1]{{#1}}

\newcommand{\daghetpart}{\algo{DagHetPart}\xspace}
\newcommand{\dagmem}{\algo{DagHetMem}\xspace}

\newcommand{\bottomlevel}[1]{\underline{l}_{#1}} % underline short italic
\newcommand{\criticalpath}{\mathcal{P}}
\newcommand{\parents}[1]{\,\Pi_{#1}}
\newcommand{\children}[1]{\,C_{#1}}
\newcommand{\cluster}{\,\mathcal{S}}

\copyrightyear{2024}
\acmYear{2024}
\setcopyright{rightsretained}
\acmConference[ICPP '24]{The 53rd International Conference on Parallel Processing}{August    12--15, 2024}{Gotland, Sweden}
\acmBooktitle{The
 53rd International Conference on Parallel Processing (ICPP '24), August
    12--15, 2024, Gotland, Sweden}
\acmDOI{10.1145/3673038.3673068}
\acmISBN{979-8-4007-1793-2/24/08}
\begin{document}

\title[Mapping Large Memory-constrained Workflows onto Heterogeneous Platforms]{Mapping Large Memory-constrained Workflows \\ onto Heterogeneous Platforms}
\titlenote{This work is partially supported by Collaborative Research Center (CRC) 1404 FONDA – Foundations of Workﬂows for Large-Scale Scientiﬁc Data
Analysis, which is funded by German Research Foundation (DFG).}

\author{Svetlana Kulagina}
\orcid{0000-0002-2108-9425}
\affiliation{%
  \institution{Humboldt Universitaet zu Berlin}
  \streetaddress{Unter den Linden 6}
  \city{Berlin}
  \country{Germany}
  \postcode{10099}
}
\email{svetlana.kulagina@hu-berlin.de}

\author{Henning Meyerhenke}
\affiliation{%
  \institution{Humboldt Universitaet zu Berlin}
  \city{Berlin}
  \country{Germany}
}
\email{meyerhenke@hu-berlin.de}

\author{Anne Benoit}
\affiliation{%
  \institution{ENS Lyon and IUF}
  \city{Lyon}
  \country{France}
}
\email{Anne.Benoit@ens-lyon.fr}
\renewcommand{\shortauthors}{Kulagina et al.}

\begin{abstract}
  Scientific workflows are often represented as directed acyclic graphs (DAGs), where vertices correspond to tasks and edges represent the dependencies between them. Since these graphs are often large in both the number of tasks and their resource requirements, it is important to schedule them efficiently on parallel or distributed compute systems. Typically, each task requires a certain amount of memory to be executed and needs to communicate data to its successor tasks. The goal is thus to execute the workflow  as fast as possible (i.e., to minimize its makespan) while satisfying the memory constraints.  %and using parallelism to speed up the execution.

  Hence, we investigate the partitioning and mapping of DAG-shaped workflows onto heterogeneous platforms where each processor can have a different speed and a different memory size. We first propose a baseline algorithm in the absence of existing memory-aware solutions. As our main contribution, we then present a four-step heuristic. Its first step is to partition the input DAG into smaller blocks with an existing DAG partitioner. The next two steps adapt the resulting blocks of the DAG to fit the processor memories and optimize for the overall makespan by further splitting and merging these blocks. Finally, we use local search via block swaps to  further improve the makespan. Our experimental evaluation on real-world and simulated workflows with up to 30,000 tasks shows that exploiting the heterogeneity with the four-step heuristic reduces the makespan by a factor of 2.44 on average (even more on large workflows), compared to the baseline that ignores heterogeneity.
  % only takes memory sizes (but not heterogeneity) into account.

\end{abstract}


%\ccsdesc[500]{Theory of computation~Scheduling algorithms}

%\received{15 April 2024}

\maketitle

\section{Introduction}
\label{sec:intro}

%%% CONTEXT %%%
Many large scientific applications, \egc for data analysis, are nowadays often
comprised of different software components expressed as tasks within a
workflow~\cite{DBLP:journals/dbsk/LeserHDEGHKKKKK21}.
Examples for such tasks can be data acquisition, data cleansing, various data
analysis tasks, and visualization of the results~\cite{10.1371/journal.pcbi.1008770}
-- executed on each data set.
Such workflows are typically represented as directed acyclic graphs (DAGs);
the vertices in the DAG represent the task and the edges express (input/output)
dependencies between them~\cite{adhikari2019survey,liuPacitti}.
Such workflows can be very large as well as
memory- and compute-intensive, so that they exhaust the resources of a
single machine and need to be executed on a parallel or distributed system.

%%% MOTIVATION %%%
For an efficient execution, the workflow should be mapped to the system
appropriately; one must decide which task is executed on which processor.
This requires to partition the workflow into smaller sub-DAGs and to assign
each sub-DAG to one processor. When doing so, a common objective function to
minimize is the makespan, \iec the total running time of the workflow~\cite{sinnen2007task}.
Since parallel and distributed systems such as (networks of) compute clusters
are often heterogeneous in terms of memory size and/or processor speed, these
different properties of the individual compute nodes should be taken into
account for a high-quality mapping.
%In particular, the memory size of a compute node must not be exceeded by a task, because otherwise there will be no solution}. %the execution will fail.
This leads to the following optimization problem: given a workflow in form of
a weighted DAG $G = (V, E)$ and a compute system~$\cluster$ with $k$ processors,
compute an acyclic $k'$-way partition of~$G$, with $k'\leq k$, and a mapping onto $\cluster$,
such that the execution of each sub-DAG on its assigned processor does not exceed the
memory of the processor, and the makespan is minimized.
This is an NP-hard problem, even with no memory constraints and a homogeneous platform~\cite{garey1979computers}, and hence large workflows are handled
by heuristics in practice.
While there certainly are heterogeneity-aware (\egc~\cite{alebrahim2017task}) and
memory-aware (\egc~\cite{bathie2021dynamic}) techniques,
we do not know of algorithms nor tools that \emph{simultaneously} handle large DAGs,
minimize the makespan, and respect memory constraints on the tasks.

%\vspace{-.2cm}

%%% CONTRIBUTION %%%
\subsubsection*{Contributions}
In this paper, we investigate the memory-constrained partitioning and mapping of DAG-shaped workflows onto heterogeneous
platforms, where each processor can have a different memory size and a different processor speed.
To this end, in the absence of heterogeneity-aware tools that also adhere to the memory constraints,
we first propose a baseline algorithm that produces valid solutions, building on a memory-efficient
traversal of the DAG. As our main contribution, we then present a partitioning-based heuristic. 
To deal with the difficulty of the problem, it is divided into four steps:
(i) partition the input DAG into smaller
blocks with a DAG partitioner, (ii) adapt the blocks of the DAG to fit the processor
memories, (iii) optimize for the overall makespan by further splitting and merging workflow blocks,
and (iv) block swaps to further improve the makespan.
An extensive experimental evaluation on real-world and simulated workflows shows that
(i) the heuristic scales to big workflows (we use up to 30~000 tasks), taking less than $11$~minutes on average
for them, and (ii) exploiting
the heterogeneity in the cluster with the four-step heuristic reduces the makespan by %\tochange{
a factor $2.44$
on average compared to the baseline that only takes memory sizes into account.
The improvement even reaches a factor of nearly $5$ for big workflows and a large cluster configuration.

% \todo{link to full version?}

%\vspace{-.15cm}
\subsubsection*{Outline}
% \textit{Outline.} 
We survey related work in Section~\ref{sec:related-work}.
Section~\ref{sec:model} presents the model and formally introduces
the optimization problem tackled in this paper.  Heuristics are described
in Section~\ref{sec:heuristics}, and both the experimental setting and
detailed results are presented in Section~\ref{sec:expe}.
We conclude and provide directions for future work in Section~\ref{sec:conc}.
%\todo{Maybe drop the above to save space.}
%\todo{Add zenodo link to bib entry}
%\vspace{-.05cm}

%%%%%%%%%%%%%%%%%%%%%%%%%%%%%%%%%%%%%%%%%%%%%%%%%%
\section{Related Work}
\label{sec:related-work}
The problem of executing a collection of tasks on various types of computing platforms has
received long-standing research attention~\cite{sinnen2007task}.
With the increasing number of scientific workflows that need to be executed multiple times,
this problem has gained renewed interest.
Currently, the most common way of representing a workflow is as a directed acyclic graph (DAG)~\cite{adhikari2019survey,liuPacitti},
hence having the most general representation of dependence constraints.

Large-scale workflows are to be executed on parallel or distributed platforms, and
the mapping problem consists in deciding where to execute each part of the workflow,
guided by a target objective function. Usually, the goal is to execute the entire workflow
as fast as possible, hence to minimize the {\it makespan}, which models the total execution time~\cite{sinnen2007task,brucker,lee2015resource}.

However, while performance is the objective, one must at the same time account for the constraints
given by the execution platform. In particular, since memory and I/O
become a bottleneck~\cite{jacquelin2011optimal,eyraud2015parallel},
one must make sure that each processor's local memory is large enough
to execute the part of the workflow that has been mapped on it.
For shared-memory platforms (we assume distributed memory instead),
Bathie \etal~\cite{bathie2021dynamic} propose, among others, a dynamic memory-aware DAG scheduling heuristic based on
an ILP formulation. Their optimization objective is cut-based and the overall approach is limited to
rather small workflows.

Scheduling for heterogeneous platforms has been investigated in the past already~\cite{benoit2008iso,fan2015survey,alebrahim2017task}, of course.
In particular, it has been remarked that the consideration of heterogeneity is a crucial aspect in scheduling
algorithms for cloud platforms~\cite{bittencourt2018scheduling}.
Astonishingly, despite their obvious importance for fault-free workflow execution, memory constraints are mostly ignored in the literature and also play no significant role in recent surveys~\cite{liu2020survey,adhikari2019survey}.
For scheduling tree-shaped workflows (a special
case of the DAGs considered here) with memory constraints, \new{we} showed in~\cite{KulaginaMB23mapping} that taking different memory sizes and
processor speeds into account can improve the makespan considerably. More precisely,
\new{we extended} a partitioning-based heuristic that maps trees on homogeneous platforms~\cite{gou2020partitioning} by making it heterogeneity-aware.
A similar goal is pursued by He \etal~\cite{He21} with their tree mapping algorithm.

When moving from tree-shaped workflows to the case of  general DAGs,
the relatively simple partitioning approaches above
do not work any longer. That is why we employ DAG partitioning, for which several tools exist,
among them \algo{dagP}~\cite{Herrmann19-SISC}. Techniques for partitioning general undirected graphs are
numerous~\cite{10.1145/3571808}, but in many cases not easily transferable to the DAG case.
Note that partitioning into (nearly-)balanced (and in case of DAGs: acyclic) blocks is \NP-hard both for general graphs~\cite{garey1979computers}
and for DAGs~\cite{DBLP:conf/wea/MoreiraPS17}.

We are not the first to employ DAG partitioning for scheduling.
The problem of memory minimization for series-parallel DAGs has been tackled by Kayaaslan \etal~\cite{KAYAASLAN20181}.
The authors propose an algorithm that fits parts of the workflow into memories of adequate size, but they do not aim at makespan
optimization.
Also, \"Ozkaya~\etal~\cite{ozkaya2019scalable} built a scheduler based on their partitioner \algo{dagP}.
This scheduler uses the \algo{dagP} output and then employs a list-based scheduling approach to optimize for makespan.
However, this scheduler does not take memory constraints into account, and
thus does not produce valid solutions for our target problem in general.
%it is therefore not applicable for our target problem.

\section{Model}
\label{sec:model}
%    \skug{Budget: 1,5 page}

We first describe the target applications, which are (large scientific) workflows,
in Section~\ref{sec.mod.work}.  Next, we define the execution
environment, a heterogeneous system (in terms of processor speed and memory size),
in Section~\ref{sec.mod.plat}.
We provide details about the optimization objective, which is to
minimize the makespan, in Section~\ref{sec.mod.make}.
Finally, we are ready to express the optimization problem in Section~\ref{sec.mod.pb}.
Table~\ref{tabnotation} summarizes the main notation.

\begin{table}[b!]
  \begin{center}
    \begin{tabular}{rl}
      \hline
      \textbf{Symbol} & \textbf{Meaning}  \\
      \hline
      $G = (V, E)$  & Workflow graph, set of vertices (tasks) and edges  \\
      $\parents{u}$, $\children{u}$ & Parents of a task $u$, children of a task $u$ \\
      $m_u$& Memory weight of task $u$ \\
      $w_u$   & Normalized execution time of task $u$ (makespan weight)    \\
      $c_{u,v}$   & Communication volume along the edge $(u,v)\in E$ \\
      $F$, $\mathcal{F}$ & A partitioning function and the partition it creates \\
      $V_i$ & Block number $i$\\ %\wrt~some $F$   \\
      $\cluster$, $k$   & Computing system and its number of processors   \\
      $p_j$, proc($V_i$)  & Processor number $j$, processor of block $V_i$ \\
      $M_j$, $s_j$   & Memory size and speed of processor $p_j$   \\
      $\beta$ & Bandwidth in the compute system  \\
      $\bottomlevel{u}$   & Bottom weight of task $u$   \\
      $\mu_G$, $\mu_i$ & Makespan of the entire workflow $G$ and of a block $V_i$ \\
      $\Gamma = (\mathcal{V}, \mathcal{E})$ & Quotient graph, its vertices and its edges  \\
      $r_u$, $r_{V_i}$  & Memory requirement of  task $u$ and of block $V_i$   \\
%      $r_{\max}$   & Maximum memory requirement \new{of a task} \\ %in a workflow    \\
      $\criticalpath$ & Critical path in a workflow  \\
      \hline
    \end{tabular}
  \end{center}
  \caption[]{Notation.} \label{tabnotation}
\end{table}

%\vspace{-.1cm}
\subsection{Workflow}
\label{sec.mod.work}
A workflow is modeled as a directed acyclic graph $G=(V, E)$, where $V$ is the set of vertices (tasks), and
$E$ is a set of directed edges of the form $e=(u,v)$, with $u,v\in V$, expressing precedence constraints between tasks.
Each task~$u \in V$  is performing $w_u$ operations, and it also
requires some amount of memory to be executed, denoted as~$m_u$.
Each edge $e=(u,v)$ has a cost~$c_{u,v}$ that corresponds to the size of the \new{(logical) output files} written by task~$u$ and used as input by task~$v$.

Hence, the \new{task memory requirement for the execution of a task~$u$ }
consists of the input files
(size of the files to be received from the parents),
the output files (size of the files to be sent to the children),
and the memory size~$m_u$:

% In total, a task~$u$ requires the following amount of memory for its execution:
\[
  r_u = \sum_{\new{(v,u)}\in E}c_{v,u} + \sum_{\new{(u,v)}\in E} c_{u,v} + m_u.
\]

\begin{figure}
%\vspace{-.1cm}
\centering%
\raisebox{-0.5\height}{\includegraphics[width=0.32\columnwidth,height=4.2cm]{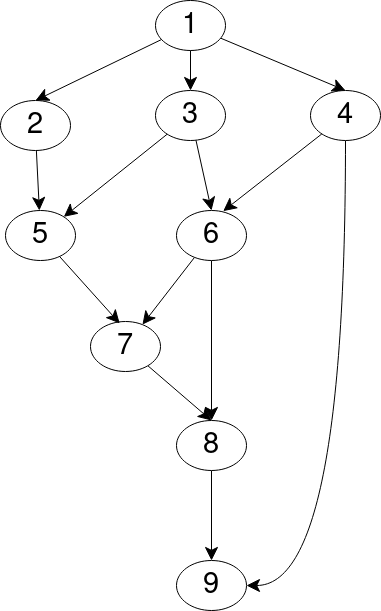}}%
\hfill%
\raisebox{-0.5\height}{\includegraphics[width=0.42\columnwidth,height=4.8cm]{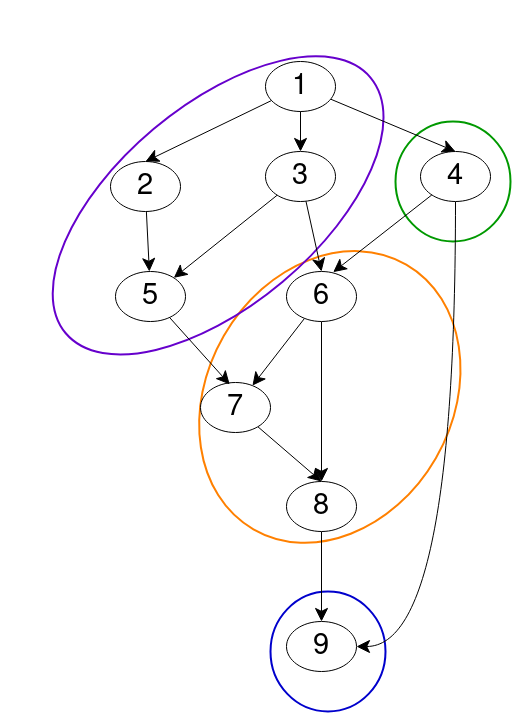}}%
\hfill%
\raisebox{-0.5\height}{\includegraphics[width=0.22\columnwidth]{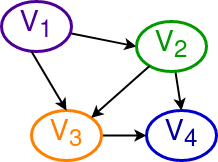}}%
 % \vspace{-.2cm}
\caption[]{An example graph $G$, its possible acyclic partition~$\mathcal{F}$ into four blocks, and a resulting quotient graph $\Gamma$.}
\Description{An example graph G, its possible acyclic partition~ mathcal F into four blocks, and a resulting quotient graph Gamma.}
  % \vspace{-.4cm}
  \label{fig:dags}
\end{figure}

%The \new{maximum} memory requirement of the workflow %(its peak memory),
%\new{when considering tasks independently is $r_{\max} = \max_{u \in V} r_u$}.
% AB: not really what we call "peak", it is the maximum over each task...
The parents of a task~$u\in V$ are the directly preceding tasks that must be completed before $u$ can be started, i.e., the set of parents is
$ \parents{u} = \{v \in V: (v,u) \in E\}$. A task without parents is called a {\it source task}.
The children tasks of~$u$ are the tasks following~$u$ directly according to the precedence constraints, i.e.,
$ \children{u} = \{v \in V: (u,v) \in E\}$. A task without children is called a {\it target task}.
%Note that since the workflow is an arbitrary DAG, 
Each task may have multiple parents and children.
Fig.~\ref{fig:dags} (left)
shows an example DAG consisting of nine tasks, with one source task ($1$)
  and one target task ($9$). The parents of task $6$ are $3$ and $4$, its children tasks $7$ and $8$.
%\AB{Cannot talk of figure if its not in main paper. I think we should keep it...}

\subsection{Execution environment}
\label{sec.mod.plat}

The goal is to execute the workflow on
a heterogeneous system, % we call cluster,
denoted as $\cluster$, which
consists of $k$ processors $p_1, \dots, p_k$.
Each processor $p_j$ ($1 \leq j \leq k$) has an individual memory size $M_j$ and a speed~$s_j$.
The execution time of a single task~$u\in V$ on a processor~$p_j$ is expressed as $\frac{w_u}{s_j}$.
We assume that all processors are connected with the same bandwidth~$\beta$.

In general, more than just one task can be mapped on the same processor; hence we aim at partitioning
the tasks into blocks, where each block will be mapped on a distinct processor.
A {\it partitioning function} is a function $F: V \rightarrow \mathbb{N}$ that assigns each task a block number.
The $i$-th block, denoted as $V_i$, contains all tasks that have been assigned number $i$: $V_i = \{ u \in V: F(u) = i \}$.
$\mathcal{F}$ is the partition, \iec the set of blocks that $F$ creates.
In Fig.~\ref{fig:dags} (middle), the partition~$\mathcal{F}$, shown as colored circles,
  divides the graph into four blocks.
If block $V_i\in \mathcal{F}$ has been assigned to a processor $p_j \in \cluster$, we say that $p_j=proc(V_i)$.
Furthermore, the maximum memory requirement of a block $V_i$ is $r_{V_i}$,
and it depends on the order in which the block, which is a DAG itself,
is executed.
We use the \algo{memDag} algorithm from~\cite{KAYAASLAN20181} to compute
this memory requirement~$r_{V_i}$, by transforming  block~$V_i$ into a series-parallel graph,
and then finding the traversal that leads to the minimum memory consumption.

Note that the number of blocks has to be at most~$k$, the number of available processors.
Some processors might be left unused, hence the DAG can be partitioned into $k'\leq k$ blocks.

\subsection{Makespan computation}
\label{sec.mod.make}

In order to compute the makespan, given a DAG~$G$ and a partitioning function~$F$,
  we consider the quotient graph $\Gamma= ( \mathcal{V}, \mathcal{E})$
  induced by the partitioning function~$F$.
%A quotient DAG $\Gamma= ( \mathcal{V}, \mathcal{E} )$ can be built from a DAG $G$ and
%a  partitioning function~$F$ as follows,
%assuming that the partition does not create cycles (and hence $\Gamma$ is a DAG).
%% \skug{explain why and about cycles}
Each quotient graph vertex $\nu_i \in \mathcal{V}$ corresponds to a block  $V_i$ of the \textit{original} DAG $G$,
as illustrated in Fig.~\ref{fig:dags} (right).
The vertex weights of $\nu_i$ ($\nu_i \in \mathcal{V}$) %, $1 \leq i \leq k$,
are defined as  $w_{\nu_i} = \sum_{u \in V_i} w_u $ (total vertex weight in block $V_i$),
while edge weights $c_{\nu_i, \nu_j} = \sum_{u \in V_i, v \in V_j}c_{u,v}$
are the sum of weights of all the edges between the two corresponding blocks.
For instance, with unitary tasks and edge weights in Fig.~1,
  $w_{\nu_1}=4$, $w_{\nu_2}=1$, $w_{\nu_3}=3$, and $w_{\nu_4}=1$.
  Furthermore, the edge costs would be $1$, except for $c_{\nu_1, \nu_3} =2$.
  In this example, $\Gamma$ is a DAG. However, note that if tasks $4$ and $9$ had been merged
  in a same block, the  %}
  resulting quotient graph would have been cyclic,
  due to the edges $(4,6)$ and $(8,9)$.
  We restrict to partitions that lead to acyclic quotient graphs, otherwise it is not
  possible to orchestrate computations and the makespan is not defined, as
  both processors rely on the result of the other processor.
  Then, given an acyclic quotient graph $\Gamma$, the makespan computation builds
  on the notion of
%\AB{will edit this}
%Before expressing the makespan of a workflow~$G$, given a mapping on a cluster~$\cluster$,
%we start with a few definitions.
%
%The
    {\it bottom weight}. % of nodes in a DAG.
  The bottom weight $\bottomlevel{\nu}$ of a node $\nu\in \mathcal{V}$ is defined as:
\begin{equation}
  \label{eq.bw}
  \bottomlevel{\nu} = \begin{cases}
                        \frac{w_{\nu}}{s_{\nu}} & \text{if $\children{\nu}=\emptyset$,}\\
                        \frac{w_{\nu}}{s_{\nu}} + \max_{\nu' \in \children{\nu}} \left\{ \frac{c_{\nu,\nu'}}{\beta} + \bottomlevel{\nu'} \right\} & \text{otherwise,}
  \end{cases}
\end{equation}
where $\children{\nu}$ is the set of children of node~$\nu$, and $s_u$ is the speed of the processor
that node~$\nu$ has been assigned to.
If node~$\nu$ has not been assigned to any processor, then the speed is assumed to be~$1$.
%\hmey{Illustrate b.w. somehow in example??}
In the example, with unitary speeds and bandwidth, we have $\bottomlevel{\nu_4}=1$
  (no children), then $\bottomlevel{\nu_3}=3+1+1=5$, $\bottomlevel{\nu_2}=1+\max\{1+5, 1+1\}=7$,
  and finally $\bottomlevel{\nu_1} = 4+\max\{1+7, 2+5\}=12$.
%\hmey{Example correct with (not yet) fixed quotient DAG in Fig. 1?
%$\bottomlevel{\nu_3} = 1 +\max\{1+5, 1+1\}=6$? }\AB{Yes indeed! Svetlana, can you
%correct Fig. 1 please? Also maybe exchange V2 and V3, it will be more logical...}

We can now define the {\it makespan} of a graph $G$
  via its corresponding quotient graph~$\Gamma$, assuming that $\Gamma$ is a DAG.
The makespan of~$\Gamma$ equals the maximum of the bottom weights of its tasks:
\begin{equation}
  \label{eq:makespan-via-bottom-level}
  \mu(\Gamma) = \max_{\nu \in \mathcal{V}} \bottomlevel{\nu}.
\end{equation}
In case of a single source, this maximum is achieved on the source task.
%
%As already mentioned, if the quotient graph were cyclic, the makespan would be undefined.
Note that in a quotient DAG $\Gamma$, when the assignment to processors changes,
  the makespan weights and the critical path change as well (in general).

An unpartitioned DAG $G$ corresponds to a quotient DAG~$\Gamma$ with only one task $\nu_1$,
and it is always executed on a single processor,~$p_j$.
Therefore, its makespan is the sum of makespan weights of its vertices divided by this processor's speed:
$\mu_G = \sum_{v \in V} \frac{w_v}{s_j}$.
Because \new{all generated files reside in the memory of the same processor, communication costs can be neglected}.

The makespan is completely defined once we have a mapping of the workflow onto a computing system~$\cluster$,
and hence we can account for processor speeds. If the mapping is not complete yet, as mentioned
in the bottom weight definition, a speed of~$1$ is assumed, so that the bottom weights can still
be computed with this assumption, leading to an \textit{estimated makespan}. If some workflow tasks
have already been assigned, the corresponding speed is used for these tasks.

Finally, note that in the definition, the finishing time of block~$V_i$ is equal to the finishing time
of all the tasks within this block, since we sum up all task weights before doing the communication
and moving up to the next block.
In reality, some tasks may finish before the block finishes,
and their successors could start earlier, but we do not consider this possibility,
hence providing in fact an overestimation of the makespan.

\subsection{Problem formulation}
\label{sec.mod.pb}

We formulate the DAGP-PM problem (\textbf{DAG P}artitioning with \textbf{P}roportionate blocks and minimized \textbf{M}akespan).

\textbf{DAGP-PM problem.}
Let $\cluster = p_1, \dots , p_k$ be the execution environment consisting of $k$  processors.
Each processor~$p_j$ (for \mbox{$1 \leq j \leq k$}) has an individual memory size $M_j$.
Given a DAG $G = (V, E)$, find an acyclic $k'$-way partitioning function $F$ (one that results
in a quotient DAG) with $k'\leq k$,  its corresponding partition
$\{ V_1, \dots , V_{k'}\}$  of $V$, and a mapping of blocks $V_i$ to processors such that
\begin{align*}
  & proc(V_i) \neq proc(V_j), \quad 1 \leq i \neq j \leq k' \\
  & r_{V_i} \leq M_i,\quad \text{where} \enspace p_i = proc(V_i)  \\
  &    \text{and the induced makespan is minimized.}
\end{align*}

In Fig.~\ref{fig:dags}, with $k=4$, the problem would require each vertex in the quotient graph to be placed on a separate
processor in such a way that the overall makespan is minimized.

Note that this DAGP-PM problem is (when formulated as decision problem) NP-complete,
even without memory constraints and precedence constraints. Indeed, the problem
of scheduling independent tasks with different execution times, even on a homogeneous platform,
is well-known to be NP-complete (by reduction from 2-partition or 3-partition~\cite{brucker}).
We are facing a complex problem, and the core contribution is to design a heuristic solution method,
\daghetpart, which accounts for the heterogeneity of the target platform while respecting the memory
constraints.

\section{Heuristics}
\label{sec:heuristics}
Before describing the \daghetpart heuristic  accounting for platform heterogeneity (Section~\ref{sec.daghetpart}),
we start in Section~\ref{sec.dagmem} with a baseline heuristic, building on related work, that aims at returning
a valid mapping of the DAG. The goal is to adhere to the memory constraints, since
no existing (makespan-oriented) mapping algorithm, to the best of our knowledge, is addressing
these constraints and is able to return a valid solution to the DAGP-PM problem.

%Section~\ref{sec.dagmem} introduces this baseline heuristic \dagmem, and then
%Section~\ref{sec.daghetpart} presents the novel
%%more involved
%heuristic \daghetpart.

\subsection{Memory-aware baseline % heuristic 
-- \dagmem}
\label{sec.dagmem}

The \new{baseline heuristic} \dagmem is directly building upon the \algo{memDag} algorithm~\cite{KAYAASLAN20181}.
\algo{memDag} computes \new{a memory-optimal} traversal of the workflow graph~$G$ that minimizes the peak
memory consumption.
\new{It however does not ensure that this traversal fits into one or multiple
actual memories of the processors.
It also does not compute the resulting makespan of the traversal, and does not try to optimize it. 
%neither optimizes for nor notes the resulting makespan of the traversal.
}

\dagmem ~\new{first} sorts the processors by decreasing memory sizes.
Thus, if the peak memory is lower than the largest memory available, the whole DAG
can be mapped on a single processor as a valid mapping.
Usually, the entire workflow does not fit into the memory of a single processor;
in this case, we perform the traversal returned by \algo{memDag} and add tasks in the first
part (block~$V_1$) as long as the memory requirement of the current (largest) processor
is not exceeded. The update to the memory requirement is done each time a new task
is traversed, by adding its memory requirement, subtracting the memory requirement
of the task executed beforehand, and updating edge weights that should remain in memory.

If the traversal of a given task~$u\in V$ leads to a block whose memory requirement is exceeding the
memory of the processor, we remove $u$ from the block.
We start by creating the next block on the next available processor, starting again from a memory of zero,
and resuming the traversal from task~$u$ that initiates the new block.

Hence, the heuristic proceeds iteratively, following the order of the initial traversal returned
by \algo{memDag}, and creating a new part as necessary. It ends %successfully
when all tasks have been traversed and assigned.
Note that for some problem instances, if the platform does not have enough memory,
this heuristic may not return any solution if there are some remaining tasks
but no more processors available. This means that the workflow needs
a larger platform to be executed.

%, or it may fail if there are some remaining tasks but no more
%processors available.

Once the blocks have been generated, the quotient graph can be computed,
based on the part numbers. Note that parts are not necessarily connected graphs,
since we follow the traversal order. But given the quotient graph, it is possible
to compute the makespan achieved by this mapping, as explained in Section~\ref{sec.mod.make}.
This baseline algorithm is not trying to optimize the makespan, but
it produces a valid solution that respects the memory constraints if it is successful.
It does not exploit the parallelism in the DAG, in particular not for workflows
that fit on a single processor memory-wise.

Hence, we focus in the following on the design of a
heuristic for the DAGP-PM
problem, aiming at minimizing the makespan while respecting memory constraints.

Note that \dagmem %is not a subroutine of \daghetpart. \dagmem 
always computes the traversal of the entire workflow graph, which may take a lot of time.
\daghetpart avoids this by partitioning the workflow graph first.

\subsection{Partitioning-based heuristic -- \daghetpart}
\label{sec.daghetpart}

The \daghetpart heuristic is going beyond \dagmem\ and aims at minimizing the makespan.
Since there are many constraints, the key idea is to decompose the heuristic
into several steps; each step builds upon the previous one to improve the solution
and adapt it to the heterogeneity of the computing system.

\subsubsection*{High-level overview}
In the first step, we use \new{a previously introduced} DAG partitioning algorithm to obtain the initial blocks that will
be used by our heuristic (hence its name \daghetpart \new{, \textbf{DAG}-shaped workflow \textbf{HET}erogeneous environment
\textbf{PART}itioning-based}).
This step ignores all the heterogeneity (memory size, processor speed) of the computing system
and optimizes for the smallest edge cut.
In the second step,  we create a preliminary assignment of these first blocks to processors, changing their size
if necessary to fit them into the memories of given processors.
This partial solution has to be valid for the assigned blocks, but there may still be unassigned blocks left.
This step respects the memory heterogeneity, but ignores the different processor speeds.
The third step uses the processor speeds to optimize the makespan. Furthermore, it aims at
respecting the number of processors, by merging %it merges
blocks if there are more blocks than processors, until it obtains a valid assignment of
blocks to processors.
Here, the algorithm may not be able to find a valid assignment for at least one block,
if the platform does not have enough resources for the DAG. In that case, no valid assignment is returned.
Finally, the fourth step starts with a valid assignment from Step~3 and tries to improve it by local search.
Two blocks are swapped (between two processors) if the operation is possible memory-wise
and if it brings the biggest improvement in makespan
among all other possible swaps.
If there are free processors, we also try to move blocks to one of these processors if it improves the makespan.
In the worst case, no improvement over Step 3 is achieved.
%\new{In this step, no further failure is possible - in the worst case, no improvement over Step 3 is achieved.}
%
We now detail each of the four steps.

\subsubsection*{\textbf{Step 1: Partitioning}}
We use the edge-cut-optimizing acyclic graph partitioner \algo{dagP}~\cite{Herrmann19-SISC} for the first step.
\new{Its input is a DAG (our workflow $G$) and a number of blocks to partition it into}.
%, which is the workflow~$G$ that we aim
%at mapping on the computing system~$\cluster$. We also specify the number of parts, \iec
%in how many blocks we want to partition~$G$.
%Since the target computing system has $k$ processors,
Computing an initial partition into $k$ blocks with \algo{dagP}
returns an array of block numbers.
For instance, with $n=5$ tasks and $k=2$ processors, we may obtain a result [1,1,1,2,2],
meaning that tasks 1,2 and 3 are in the same block~$V_1$, and tasks 4 and 5 are
in another block~$V_2$.
Note that we tentatively partition the DAG into $k'$ blocks, with $1\leq k'\leq k$,
and compute the makespan returned by the heuristic for all values of~$k'$. 
The best result is kept.

%\vspace{-.1cm}
\subsubsection*{\textbf{Step 2: Assignment}}
We start with the partitioning function obtained from Step~1, and we
insert the resulting blocks into a priority queue~$Q$ -- with the required block memory size as priority.
The processors, in turn, are inserted into a queue~$M$ in descending
order \wrt
their available memory size (see Algorithm~\ref{alg:BiggestAssign}).
Then, we assign the largest block to the free processor with the largest memory
as long as there are remaining blocks and processors (while loop starting in Line~\ref{line:whilebig}).
During this assignment (algorithm~\algo{FitBlock}), it may happen that the block does not fit on the processor, \iec its memory requirement
exceeds the memory available on the processor. In this case the block is further
partitioned, and the resulting sub-blocks are inserted back into the queue~$Q$.
Hence, the size of $Q$ can become bigger than that of~$M$.
If this happens, the
unassigned blocks are further partitioned (loop of Line~\ref{line:remain-unassigned})
to the size of the memory of the smallest processor in the {computing system} -- by another call
to \algo{FitBlock} that will not perform any assignment.

\begin{algorithm}
  \caption[]{-- Step 2: \algo{BiggestAssign}}
  \label{alg:BiggestAssign}
  \begin{algorithmic}[1]
    \Procedure{\algo{BiggestAssign}}{$\mathcal{F}$, $\cluster$}\\
    \Comment{Input: initial partition $\mathcal{F}$, {computing system} $\cluster$.}
    \State \texttt{Init} PQ $Q$ with $\mathcal{F}$;\label{line:pq}
    \Comment{max-priority queue of the blocks}
    \State  \texttt{Init} queue $M$ with $\cluster$;\label{line:pq2}\\
    \Comment{a queue of processors sorted by memory size}
    \While{\textbf{not} $Q$\texttt{.empty()} \textbf{and not} $M$\texttt{.empty()}}
      \label{line:whilebig}
      \State $V_{m} \gets Q$\texttt{.extractMax()};  $p_{m} \gets M$\texttt{.head()};
      \State $V_r \gets $\algo{FitBlock}($V_{m}$, $Q$, $p_{m}$, $true$); \label{line:fit-block} \\
      \Comment Fit the block while actually trying to map it
      \If{$V_r \neq$ NULL}
        $M$\texttt{.remove}($p_m$); \\
        \Comment{Remove the processor that is now busy}
      \EndIf
    \EndWhile
    \State $p_{\min} \gets \arg\min M(\cluster)$;\label{line:smallest-proc}
    \Comment{processor with smallest memory}
    \While{\textbf{not} $Q$\texttt{.empty()}}
      \label{line:remain-unassigned}
      \Comment{No more processors, but remaining partition blocks}
      \State $V_{m} \gets Q$\texttt{.extractMax()};
      \State \algo{FitBlock}($V_{m}$, $Q$, $p_{\min}$, $false$); \label{line:fit-to-smallest}\\
      \Comment Fit the block without actually putting it on the proc.
    \EndWhile

    \EndProcedure
  \end{algorithmic}
  \end{algorithm}

The subroutine \algo{FitBlock} (Algorithm~\ref{alg:FitBlock}) works as follows.
Its input is the block~$V_i$ to be fitted, the priority queue of blocks, the processor where the block
should be scheduled on, and a flag \texttt{doMap} that indicates that we really want to place the block on this processor -- rather than just partitioning it down to the processor's memory size.
If a block $V_i$ fits on the desired processor and if the flag is true,
the block is mapped on its target processor % (Line~\ref{line:schedule-if-fits})
and the algorithm returns the block that was placed.
Otherwise, we further partition this block using the same partitioner as
in Step 1. % (Line~\ref{line:repartition} in Algorithm~\ref{alg:FitBlock}).

\new{
\begin{algorithm}[h]
\caption[]{-- Step 2 subroutine: \algo{FitBlock}}
\label{alg:FitBlock}
\begin{algorithmic}[1]
\Procedure{FitBlock}{$V_i$, $Q$, $p_j$, \texttt{doMap}}
\Comment{Input: a block $V_i$ that needs to be fitted, priority queue $Q$ of other blocks,
available proc. $p_j$, a flag \texttt{doMap} that indicates that we want to map the resulting subtree on the processor}
\If{$r_{V_i} \leq M_{j}$}
\If{\texttt{doMap}}
\State proc($V_i$) $\gets p_j$;\label{line:schedule-if-fits}
\State \textbf{return} $V_i$;
\EndIf
\Else
\State $(V_{m_1}, V_{m_2},...) \gets $ \algo{Partition}($V_m$, 2); \label{line:repartition}
\For{$V_k \in \{V_{i_1}, V_{i_2},...\}$}
\State  $Q$\texttt{.add}($V_k$);  \Comment{Reinsert resulting blocks}
\EndFor
\EndIf
\State \textbf{return} NULL;
\EndProcedure
\end{algorithmic}
\end{algorithm}

}
Although we strive for dividing the block into two new blocks, sometimes \algo{dagP} cannot
do that due to balancing constraints. Hence, \algo{Partition} may generate more than two blocks.
We then insert the resulting blocks back into the queue~$Q$ and a NULL result is returned,
indicating that no block has been placed on the processor.
Because $Q$ is a priority queue, it maintains the correct order of the blocks in terms of their memory requirement.
Otherwise, a relatively small block obtained during repartitioning could block a bigger waiting block from
being assigned.
%describes the fitting procedure in detail.

Step 2 returns at least a valid partial assignment of blocks to processors, but not necessarily a complete one:
all assigned blocks have to fit, but some blocks may not be assigned yet.

%\vspace{-.1cm}
\subsubsection*{\textbf{Step 3: Merging}}
The focus of this step lies in the makespan, affected by heterogeneous processor speeds.
We operate on the quotient graph $\Gamma$ that is built from the original graph $G$ using the blocks from Step 2 and
their (partial) assignment to processors.
According to the definition of the quotient graph, each block~$V_i$ of the original graph becomes
a vertex in~$\Gamma$.
Since the assignment in Step 2 may have been incomplete, some of these vertices are assigned to processors, while some may not.
We try to improve the makespan by merging unassigned vertices in~$\Gamma$
to the assigned ones in a way that
impairs the makespan the least.
Because there are still unassigned vertices, we operate with the \textit{estimated makespan}. % for this step.

We first observe that merging two unassigned vertices in~$\Gamma$ will not help our cause:
such a merge results in a bigger,
still unassigned vertex that will be more difficult to place on a processor.
Therefore, we merge an unassigned vertex with an assigned one.

The next observation is that the makespan is constrained by the critical path in~$\Gamma$.
Indeed, the critical path is the one that dictates the makespan when computing
the bottom weights (see Eq.~\eqref{eq.bw}).
Merging more vertices to one that lies on the critical path will increase the makespan,
while merging outside of the critical path will
only affect the makespan if the merge changes the critical path.
Hence, we prefer merging an unassigned vertex outside of the critical path to a vertex
that is not on the critical path, but we still allow merges on the critical path
if no other solution is possible. In particular, if a node on the critical path is not assigned
and considered for merging,
it can be interesting to merge it to its parent or children, thus saving communication % costs
on the critical path.

\begin{figure}[b]
  \centering
  \begin{tikzpicture}[xscale=.7,yscale=.5]
    % Vertices
    \node[circle, draw, minimum size=0.45cm, fill=cyan!20!white] (A) at (0,0) {$a$};
    \node[circle, draw, minimum size=0.45cm, fill=cyan!20!white] (B) at (0,-2) {$b$};
    \node[circle, draw, minimum size=0.45cm, fill=cyan!20!white] (C) at (1.5,-1) {$c$};

    \node[circle, draw, minimum size=0.57cm, fill=cyan!20!white] (AB) at (3.5,-1) {$a\!+\!b$};
    \node[circle, draw, minimum size=0.57cm, fill=cyan!20!white] (CC) at (5.5,-1) {$c$};

    \node[circle, draw, minimum size=0.57cm, fill=cyan!20!white] (ABC) at (8,-1) {$a\!+\!b\!+\!c$};

    % Arrows
    \draw[->] (A) -- (B);
    \draw[->] (A) -- (C);
    \draw[->] (C) -- (B);
    \draw[->,bend left=30] (AB) to (CC);
    \draw[->,bend left=30] (CC) to (AB);
  \end{tikzpicture}

 % \vspace{-.1cm}
  \caption[]{Merging two vertices %in an acyclic graph 
  can create a cycle of length~2. Merging all three vertices may solve the problem.}
  \Description{Merging two vertices %in an acyclic graph
can create a cycle of length~2. Merging all three vertices may solve the problem.}
  \label{fig:mergecyc}
 % \vspace{-.2cm}
\end{figure}
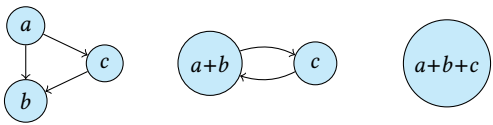

First, let us consider an algorithm that finds the best possible merge for a single quotient
vertex $\nu$~(Algorithm~\ref{alg:MsOptMerge}).
It takes the vertex $\nu$ in question, a list of candidates that can be used for merging~$A$, and the graph $\Gamma$ as the input.
We can merge~$\nu$ to one of either its parents or its children, but only those that are in $A$~(Line~\ref{line:merge-to-who}).
For each such vertex $\nu'$, we tentatively merge it with $\nu$, an action that results in the creation of a new
vertex $\nu_m$, and changes $\Gamma$ to $\Gamma'$~(Line~\ref{line:tentat-merge}).
In the bad case, $\Gamma'$ contains cycles after this action, as in Fig.~\ref{fig:mergecyc}
if $a$ and $b$ are merged.
In the general case, introducing a cycle into the graph is a reason to discard the tentative merge.
However, there is one easily solvable case, illustrated by Fig.~\ref{fig:mergecyc}.
In this case, the merge creates a cycle of length~2.
This situation can (often) be resolved by merging the third vertex into the set.

This is exactly what we do if we encounter a cycle of length~2 (Line~\ref{line:ifcyc}): we {tentatively} merge the already merged
vertex~$\nu_m$ with the second vertex in the cycle (Line~\ref{line:merge3}).
If the graph still has cycles after that, then we discard this merge: we restore $\nu_m$ to its original state and continue with another
merge candidate for $\nu$.
Otherwise, we store the third vertex as $\nu_o$, an optional third vertex,
and operate on the graph $\Gamma''$ that
contains this triple-merged vertex.

If $\Gamma'$ remains cycle-free or if a single $2$-vertex cycle has been eliminated, we assess the quality of this
merge (Line~\ref{line:merge-fits}).
The memory requirement of $\nu_m$ must not exceed the memory available on the processor of $\nu'$, the merge candidate.
We compute the makespan of the modified~$\Gamma'$ (Line~\ref{line:assess-makespan}).
If this is the best makespan we have encountered yet, we store it and the vertex that allowed it
(Line~\ref{line:ifbestms})
and proceed to try out the next merge.
We return the best makespan found, the vertex that yielded it, and the optional third vertex.

\begin{algorithm}[tb]
  \caption[]{-- Step 3 subroutine: \algo{FindMSOptMerge}}
  \label{alg:MsOptMerge}
  \begin{algorithmic}[1]
    \Procedure{FindMSOptMerge}{$\nu$, $A$, $\Gamma$}
      \Comment A block $\nu$ (vertex in~$\Gamma$)  that needs to be merged, a set $A$ of vertices that can be used for the merge, the graph $\Gamma$.
      \State $\mu_{min} \gets +\infty$; $\nu_{min} \gets NULL$; $\nu_o \gets NULL$;\\
      \Comment $\nu_o$ is an optional third vertex in the merge
      \For{\textbf{each} {$\nu' \in (\parents{\nu} \cup \children{\nu} ) \cap A $}}
        \label{line:merge-to-who}
        \State $\{ \nu_m, \Gamma'\} \gets$ \algo{Merge}($\nu$,$\nu'$);
        \label{line:tentat-merge}
        \If{isCyclic($\Gamma'$) }
          \If{Cycle($\Gamma'$).size=2}
            \label{line:ifcyc}
            \State \{$\nu_m, \Gamma'' \} \gets$ \algo{Merge}($\nu_m$,Cycle($\Gamma'$)[1]);\label{line:merge3}
            \If{isCyclic($\Gamma''$) }   \algo{Unmerge}($\nu_m$); break;
            \Else $\quad\nu_o \gets$ Cycle($\Gamma$)$[1]$; $\Gamma' \gets \Gamma''$;
            \EndIf
          \Else $\quad$  \algo{Unmerge}($\nu_m$); break;
          \EndIf
        \EndIf
        \If{$r_{\nu_m} \leq M_{proc(\nu')}$}
          \label{line:merge-fits}
          \State $\mu \gets$ \algo{makespan}($\Gamma'$); \label{line:assess-makespan}
          \Comment Assess the impact of this merge on the makespan of the entire graph
          \If{$\mu \leq \mu_{min}$}
            \label{line:ifbestms}
            \State $\mu_{min} \gets \mu; \nu_{min} \gets \nu'$;
            \State $\Gamma \gets \Gamma'$;
          \EndIf
        \EndIf
        \State \algo{Unmerge}($\nu_m$);
      \EndFor
      \State \textbf{return} ($\mu_{min}, \nu_{min}, \nu_o $);
    \EndProcedure
  \end{algorithmic}
 % \vspace{-1ex}
\end{algorithm}

\begin{algorithm}
  \caption[]{-- Step 3: \algo{MergeUnassignedToAssigned}}
  \label{alg:Merge}
  \begin{algorithmic}[1]
    \Procedure{MergeUnassignedToAssigned}{\{$V_1, \ldots, V_k$\}, $P$} $\quad$
      \Comment{Input: Blocks $V_1, \ldots, V_k$, processors $P$. }
      \State $\Gamma \gets $ \algo{QuotientDAG}($G$, \{$V_1, \ldots, V_k$\} ); \Comment Quotient DAG from the blocks of $G$ \label{line:gamma}
      \State $\criticalpath \gets $ \algo{CriticalPath}($\Gamma$);
      \State $A \gets \{ \nu \in \mathcal{V} : proc(\nu) \neq NULL\}$; \label{line:set-A}
      \Comment Assigned vertices
      \State $U \gets \{ \nu \in V(\Gamma)  \!:\! proc(\nu) = NULL$ \};  \label{line:set-U}
      \Comment Unassigned vertices
      \For{$\nu \in U$}
        \label{line:iter-unassigned}
        \State $U \gets U \setminus {\nu}$; $\mu_{min} \gets \infty$; $\nu_{min} \gets NULL$;
        \State  $\mu_{min},\nu_{min}, \nu_o \gets $ \algo{FindMSOptMerge}($\nu$,$A \setminus \criticalpath $,$\Gamma$);\label{line:notoncp}
        \If{$\nu_{min} \neq NULL$}
          \label{line:found-merge}\Comment Found a feasible merge
          \State \{ $\nu_m, \Gamma' \} \gets$ \algo{Merge}($\nu$,$\nu_{min}$,$\nu_o$); \label{line:execute-merge}
          \State $\Gamma \gets \Gamma'$; $A.$remove($\nu_{min}$); $A.$remove($\nu_{o}$);
          \State proc($\nu_m$) $\gets$ proc($\nu_{min}$);
          \State $\criticalpath \gets $ \algo{CriticalPath}($\Gamma$);
          \Comment Critical path could have changed after the merge.
        \Else
          \Comment No feasible merge in $A \setminus \criticalpath$
          \State  $\mu_{min},\nu_{min}, \nu_o \gets $ \algo{FindMSOptMerge}($\nu$,$A$,$\Gamma$); \label{line:oncp}
          \Comment Look for a feasible merge anywhere in $\Gamma$
          \If{$\nu_{min} \neq NULL$}
            \Comment Found a feasible merge %somewhere
            \State \{$\nu_m, \Gamma' \} \gets$ \algo{Merge}($\nu$,$\nu_{min}$,$\nu_o$); \label{line:execute-merge1}
            \State $\Gamma \gets \Gamma'$; $A.$remove($\nu_{min}$); $A.$remove($\nu_{o}$);
            \State proc($\nu_m$) $\gets$ proc($\nu_{min}$);
            \State $\criticalpath \gets $ \algo{CriticalPath}($\Gamma$);
          \Else
            \If{($\children{\nu} \cap U \neq \emptyset$ \textbf{or} $\parents{\nu} \cap U \neq \emptyset$ )
              \textbf{and} $\nu.c \leq 1$}
                 %\State 
              $U \gets U \cup \{\nu\}$; $\nu.c++$;
              \Comment Unassigned parents or children
              \label{line:unassignedparents}
         \EndIf
          \EndIf
          \State \textbf{return} false; \label{line:fail}
          \Comment No solution could be found %Failure due to not finding a fit

        \EndIf

      \EndFor
      \State \textbf{return} $\Gamma$;
    \EndProcedure
  \end{algorithmic}
  %\vspace{-1ex}
\end{algorithm}

With Algorithm~\ref{alg:MsOptMerge} described, we can now formulate the merging algorithm
used in Step~3 (Algorithm~\ref{alg:Merge}).
We first construct a quotient DAG $\Gamma$~(Line~\ref{line:gamma})
and compute its critical path~$\criticalpath$.
Each vertex in $\Gamma$ additionally has a counter~$z$ (number of times it was seen) that is set
to $0$ at the beginning.
Then, we build the set $A$ of all assigned tasks (Line~\ref{line:set-A}).
We iterate over all unassigned blocks (Line~\ref{line:iter-unassigned}).
When doing so, we first try to merge the unassigned block to an assigned one that is not
on the critical path (Line~\ref{line:notoncp}).
If a feasible merge has been found this way, we execute it in Line~\ref{line:execute-merge},
then assign the resulting
vertex to the processor, and recompute the critical path.
Otherwise, we repeat the same attempt, but now we try to merge to any assigned task -- no matter if it is on the critical
path or not~(Line~\ref{line:oncp}).
If this merge succeeds, we execute it in the same fashion.
If no merge {can be} found even on the critical path, then we consider the possibility of trying to merge this vertex
at a later point in time.
If a vertex has unassigned parents or children, it may become mergeable later, when these parents or children have
themselves been assigned~(Line~\ref{line:unassignedparents}).
In this case, we reinsert the vertex into the set of unassigned vertices to be checked later.
However, to avoid a situation where two vertices are being constantly reinserted after each other, we use the counter $\nu.c$.
We increase it on each reinsert and stop reinserting after 2 times.
Finally, we return the quotient graph (with the assigned tasks). % if we succeed.

%\vspace{-.1cm}
\subsubsection*{\textbf{Step 4: Swaps}}
%
 % in extended version~\cite{daghetpart_full_version})}, 
We perform local search via block swaps to further improve the makespan.
We first identify all feasible swaps (pairs of vertices that fit into the respective memory 
of the processor of the other vertex). 
%in Lines~\ref{line:forall-pairs}-\ref{line:feasible-swaps}.
The best swap is stored %~(Line~\ref{line:improvement-found}) 
and executed.
We stop when no more makespan-improving swaps can be made
(see Algorithm~\ref{alg:SwapUntilBest}).

\begin{algorithm}[htb]
\caption[]{-- Step 4: \algo{Swap} (swap best pair % \algo{ Phase A: Swap the best pair
until no feasible improving swap exists)}
\label{alg:SwapUntilBest}
\begin{algorithmic}[1]
\Function{\algo{Swap}}{$\Gamma$, $\cluster$}
\Comment Quotient DAG $\Gamma$ whose tasks $\mathcal{V}$ have been assigned to processors, {computing system} $\cluster$
\State $best \gets$ \texttt{pair}($\Gamma$, \texttt{makespan}$(\Gamma)$); \label{line:init-best}
\While{true}
\State $curr \gets best$;
\ForAll{pairs $(\nu, \nu'')$ of $\mathcal{V} \times \mathcal{V}$}
\label{line:forall-pairs}
\If{\texttt{isSwapFeasible}($best.first$,$\nu$,$\nu'$)}
\label{line:feasible-swaps}
\State $\Gamma' \gets$ \texttt{swap}$(best.first, (\nu, \nu'))$; \label{line:swap}
\State $next \gets$ \texttt{pair}($\Gamma'$, \algo{makespan}$(\Gamma')$);
\If{$next.second < curr.second$}
\State $curr \gets next$; \label{line:improvement-found}
\Comment{better solution is stored}
\EndIf
\EndIf
\EndFor \label{line:forall-pairs-end}
\If{$curr.second < best.second$}
\State $best \gets curr$; $\Gamma \gets  best.first$;  \label{line:new-best}\\
\Comment{execute improving swap and work further with the resulting quotient graph}
\Else
~\textbf{return} $best$; \label{line:return}\\
\Comment{stop because no improving swap exists}
\EndIf
\EndWhile
\EndFunction
\end{algorithmic}
\end{algorithm}

Furthermore, it may happen, in particular with small workflows,  that
the partitioner does not create enough blocks and some of the processors
remain idle. In that case, some of these idle processors may be faster
than those the blocks were assigned to.
For this reason, we add one final step that is only activated if there are idle processors after swapping.
In this step, we go through the critical path of $\Gamma$ and try to move each task to a faster idle processor that can hold it memory-wise.
If such a processor exists, we move the task there and recompute the critical path.
We do this as long as there are tasks in the critical path that have not been moved.

\section{Experimental Evaluation}
\label{sec:expe}
%    \skug{budget: 3 pages}
%Section~\ref{sec:setup} describes the setup for evaluating the \daghetpart heuristic,
%Section~\ref{sec:results} presents the results.

\subsection{Setup}
\label{sec:setup}

All algorithms are implemented in C++ and compiled with g++ (v.11.2.0).
\new{\dagmem, proposed in Section~\ref{sec.dagmem} as a variant of the existing \algo{memDag}, functions as
baseline algorithm -- because there is no previous work in our scenario that respects the memory constraints.}
Exceeding the memory would yield invalid and thus incomparable mappings.
The experiments are executed on workstations with 192 GB RAM and 2x 12-Core Intel Xeon 6126 @3.2 GHz
and CentOS 8 as OS.
Code, input data, and experiment scripts are available for review at \url{https://zenodo.org/records/10992671}.
%together with an extended 
%version of the paper containing an appendix with additional results~\cite{daghetpart_full_version}.}
%can be downloaded at \url{https://zenodo.org/records/10569313}.

We first describe  the set of workflows used in the evaluation and then the clusters on which the
workflows are mapped.

\subsubsection{Workflow instances}

The input data for the experiments consists of two sets of workflows: real-world workflows
obtained from~\cite{ewels2020nf} and workflows obtained by simulating real-world workflows
with the WFGen generator~\cite{COLEMAN202216}.
First, we discuss how the graph topology is generated,
then we focus on the weights associated to tasks and edges.

%\vspace{-.1cm}
\paragraph{Workflow graphs}
For real-world workflows, their %\href{https://github.com/nextflow-io/}{
nextflow definition (see~\cite{ewels2020nf}) was downloaded from the respective repository
and transformed into
.dot format using the nextflow option ``-with-dag''.
The resulting DAG contains many pseudo-tasks that are only internal representations in nextflow
(and not actual tasks); that is why we removed them.

For the simulated workflows, the graph is produced by the WFGen generator,
based on a {\em model workflow} and the desired
number of tasks.
The model workflows are described on the WFGen website, and we used the following ones:
1000Genome, BLAST, BWA,
Epigenomics, Montage, Seismology, and SoyKB.
Other models could not be generated without errors.
As number of tasks, we use: 200, {1~000}, {2~000}, {4~000}, {8~000}, {10~000},
{15~000}, {18~000}, {20~000}, {25~000}, {30~000}.
We divide the workflows into three groups by size: small ones with up to {8~000} tasks, middle
ones with {10~000} to {18~000} tasks, and big ones with {20~000} to {30~000} tasks.
{Note that for some workflow models, such as SoyKB or Montage, only a subset of the workflow sizes could be generated, because the workflow generator
either took a disproportionally long time or yielded errors.}

Overall, this yields four workflow types, denoted by real, small, mid \new{(middle-sized)}, and big.

%\vspace{-.1cm}
\paragraph{Generation of task and edge weights}
For the real-world workflows, we use historical data files provided by Bader~\etal~\cite{lotaru}.
The columns in these files are measured Linux PS stats, acquired during an execution of a nextflow workflow.
Each row corresponds to an execution of one task on one cluster node.
Since the operating system cannot distinguish between (a) the RAM the task uses for itself and (b) the RAM it uses
to store files that were sent or received from other tasks, the values in the historical data are \new{task memory requirements} (input/output files plus memory consumption of the computation).
In a similar manner, the historical data provided by~\cite{lotaru} do not store the actual weights of edges between tasks, but only the overall
size of files that the task sends to all its children.

For each task, historical data can contain multiple values, obtained from the runs with different input sizes.
To avoid underestimation, we take the maximum among all values from different runs on the same cluster node.
Not all tasks have historical runtime data stored in the tables.
In fact, for two workflows, Bader~\etal do not provide data for more than 50\% of the tasks.
For two more, around 40\% of tasks have no historical runtime data stored.
Hence, in the absence of historical data about a task, we give it a weight of~1.
Because the historical data contains absolute measured values and the cluster node information is a relative value,
we normalize all values extracted from the historical data by the smallest one.
This way, task memory weights fit into the memory of the cluster nodes.
Additionally, this way, the tasks without historical data receive less insignificant values compared to tasks with historical data.

For the simulated workflows, we generate random execution times and memory weights for each task as well as
edge weights for each precedence constraint.
We generate uniformly distributed values between $1$ and $10$ for edge weights,
$1$ and $1000$ for the workloads,
and $1$ and $192$ for memory weights.
When doing so, we try to mimic the weights observed in the historical data, hence \eg the low lower bounds for the
workloads.

\subsubsection{Target computing systems}
To fully benefit from the historical data, the  {\em default} experimental environment
that we consider is a cluster based on the same six
kinds of real-world machines that were used in the experimental evaluation in~\cite{lotaru}.
We set the number of each kind of node to six, thus having 36 processors in total. % the whole cluster.

Each machine has a (normalized) CPU speed and a memory size (in GB), and we list them as (name, speed, memory): 
($local$, 4, 16) -- very slow machines; ($A1$, 32, 32), ($A2$, 6, 64), ($N1$, 12, 16) -- average machines; ($N2$, 8, 8) -- machine with very small memory;
and ($C2$, 32, 192) -- {\em luxury} machine with high speed and large memory
(see Table~\ref{tab:procs}).

\begin{table}[htb]
\begin{center}
\begin{tabular}{rcc}
\toprule
Processor name &  CPU speed (GHz)  &Memory size (GB) \\
\midrule
local                    & 4                    & 16     \\
A1                      & 32                   & 32     \\
A2                      & 6                    & 64     \\
N1                      & 12                   & 16     \\
N2                      & 8                    & 8      \\
C2                      & 32                   & 192    \\
\bottomrule
\end{tabular}
\end{center}
\caption[]{Cluster configuration (default).}
\label{tab:procs}
\end{table}

Note that, by nature, memory sizes are normalized values, relative to each other, too.
Therefore, we additionally normalize memory weights of real-world workflows
to the maximum size of 192 to make sure they fit.
For simulated workflows, we increase memory sizes proportionally until the task
with the biggest memory requirement still
has a processor it could be executed on.

To test various settings, we also vary the cluster configuration and consider
variants of the default cluster presented above:

\noindent  $\bullet$ {\em Small} and {\em large} clusters. While the default cluster consists of 36 nodes (6 of each kind), we
  consider a small cluster with three processors of each kind ($18$ processors in total), and a
  large cluster with ten processors of each kind ($60$ processors in total).

\noindent  
$\bullet$ More and less heterogeneous clusters ({\em MoreHet}, {\em LessHet}, {and {\em NoHet}}).
  In addition to the default cluster, we also consider more and less
  heterogeneous clusters (Table~\ref{tab:procshetero}).
  For a cluster with more heterogeneity, the smaller half of memories is made twice smaller, whereas
  the bigger half is made twice bigger.  The same holds for processor speeds.
  For the less heterogeneous cluster, we reverse the procedure: smaller values are increased two-fold,
  while bigger values are reduced. An exception is made for the biggest memory size: we leave it at 192 to
  make sure that the largest memory requirements of tasks can still be met.
  {In the homogeneous cluster {\em NoHet}, each processor has to be able to hold even the most memory-demanding task.
  Therefore, in this case, all processors have to be~$C2$, the processor with the biggest memory capacity.}
 % \item %Communication to computation costs.

\noindent $\bullet$ 
  Finally, it is interesting to examine the impact of the commu\-nication-to-computation ratio  ({\em CCR})
  by modifying the bandwidth~$\beta$ (also used in the makespan computation).
  To explore different scenarios, we vary $\beta$ between $0.1$ and $5$.

  \begin{table}[h!]
  \begin{center}
    \begin{tabular}{rcc|rcc}
      \toprule
      {\em MoreHet} &   Speed  &Memory  & {\em LessHet} &  Speed &Memory \\
      \midrule
      local*                   & 2                    & 8    &  local'                 & 8                    & 64   \\
      A1*                     & 64                   & 64   & A1'                   & 16                   & 64    \\
      A2*                     & 3                    & 128   &A2'                   & 12                    & 128    \\
      N1*                      & 24                   & 8    &N1'                    & 12                   & 64    \\
      N2*                     & 4                    & 4      &N2'                   & 16                    & 32 \\
      C2*                   & 64                   & 384   & C2'                   & 16                   & 192  \\
      \bottomrule
    \end{tabular}
  \end{center}
  \caption[]{Clusters with more or less heterogeneity.\\ {Memory size is in GB, speed is in GHz.}}
  \label{tab:procshetero}
  \end{table}

%\end{itemize}

\subsection{Results}
\label{sec:results}

We first present results on the default cluster for all types of workflows. 
We then study the impact of other cluster configurations, report running times, and summarize the results.

\subsubsection{Default cluster}
\begin{figure}
\centering
\includegraphics[width=0.53\columnwidth]{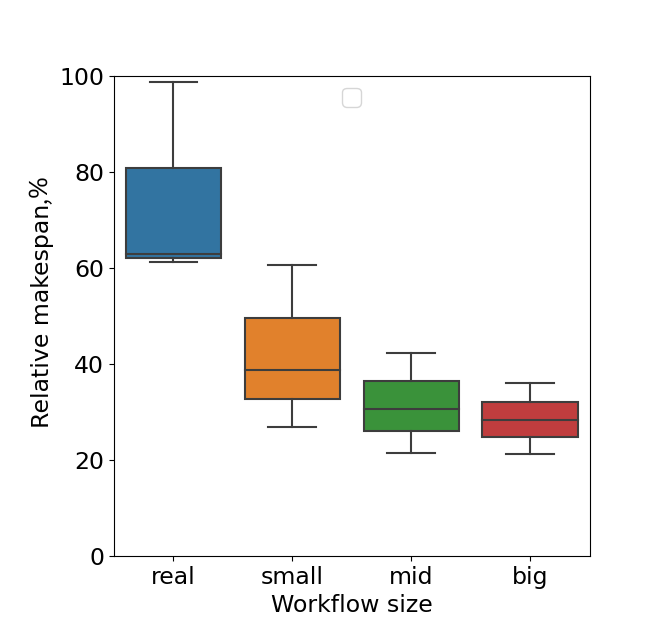}%
\hfill%
\includegraphics[width=0.46\columnwidth]{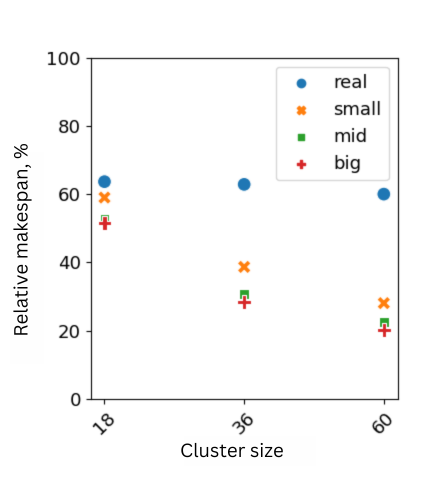}%
%\vspace{-.5cm}
\caption[]{Left: Relative makespan (in \%) of \daghetpart compared to \dagmem on default cluster.
  Right: Relative makespan (in \%) on different cluster sizes ($x$-axis: number of CPUs), by workflow size. Smaller is better. }
  \Description{Left: Relative makespan (in \%) of \daghetpart compared to \dagmem on default cluster.
Right: Relative makespan (in \%) on different cluster sizes ($x$-axis: number of CPUs), by workflow size. Smaller is better.}
  \label{fig:finems} \label{fig:improvementbysize}
  %\vspace{-0.3cm}
\end{figure}

Fig.~\ref{fig:finems} (left) shows the relative makespan of \daghetpart\
(ratio of makespans by \daghetpart and \dagmem, in \%),
by workflow types (geometric mean over the ratios of each workflow); hence the lower, the better.
% For instance, 25\% means that the achieved makespan is four times lower than the one of the baseline.
When averaging over all workflow types, the relative makespan is 41\%;
hence \daghetpart yields mappings that are on average $2.44\times$ better in makespan than the baseline.
The trend is that \daghetpart can achieve the best improvement on workflows of the categories big ($3.3\times$ better
  over all workflows, $4.82\times$ better on the most fanned-out families),
and middle ($3.23\times$ better over all, $4.84\times$ on fanned-out).
Yet, even on real-world workflows (which are small, the smallest one has only $11$ tasks),
\daghetpart is still $1.59\times$ {better}.

\new{
On the default cluster,
\daghetpart can schedule $13$ workflows out of $14$ big ones and $31$ out of $32$ small ones.
{The baseline was also unable to schedule the small workflow, but scheduled the big one.}
All middle-sized  and  all real-world workflows can be scheduled successfully {by both \daghetpart and \dagmem}.
%\todo{Anne, do you want to comment this here already (not: failure)?}

It is more challenging with real-world workflows to improve on \dagmem due to several reasons.
The first one is their small size, ranging from only~$11$ to $58$ tasks.
As a consequence, the partitioner is unable to decompose these workflows into the desired number of blocks
($36$, as the number of nodes in the cluster);
instead, the workflows are split into $11$ to $14$ blocks in the first step of \daghetpart.
This already severely limits the algorithm's ability to exploit both the parallelism and the heterogeneity.
On top of that, more than half of the tasks in real-world workflows have no historical data.
This leads to a situation where relatively few tasks have actual values and a long ``tail'' of tasks has values of $1$.
It makes sense to schedule these tiny tasks together on one machine, which also limits the ability to
use different machines in the cluster.
}
%
%Based on the results on simulated data, we consider it reasonable to assume that \daghetpart
%would work well on large real-world workflows that would come with sufficient historical
%  (or estimated) task runtime data.

\subsubsection{Impact of parallelism -- small and large clusters}
Exploiting parallelism in the workflows is the main reason for \daghetpart's success.
Being able to utilize even the slower/smaller (memory-wise) processors yields higher improvements on clusters with more nodes:
\daghetpart yields on large workflows makespans that are $4.96\times$ better.
Fig.~\ref{fig:improvementbysize} (right) illustrates this trend.
The trend is less obvious for real-world workflows, because 
only one real-world workflow is big enough to benefit from the large cluster.
The remaining four do not occupy even all
nodes in the default cluster, so that adding new ones has little effect on the makespan.

Moreover, on the large cluster, we are able to schedule all workflows from all types,
while no solutions were found for two workflows on the default cluster.
On the small cluster with $18$ nodes, we are able to schedule $13$ out of $14$ big workflows (\dagmem: $13$),
$13$ out of $17$ middle-sized workflows (\dagmem: $15$),  and
$30$ out of $32$ small workflows (\dagmem: $31$). 
The algorithms did not succeed on some configurations where 
there are not enough resources for the workflow. It is non-trivial to figure
out whether a solution would be achievable, other than doing an exhaustive search.
We believe that the user should rather consider using a larger platform in these cases.

\subsubsection{Impact of heterogeneity -- NoHet, MoreHet, and LessHet clusters}
\begin{figure}[tb]
  \centering
  \includegraphics[width=0.495\columnwidth]{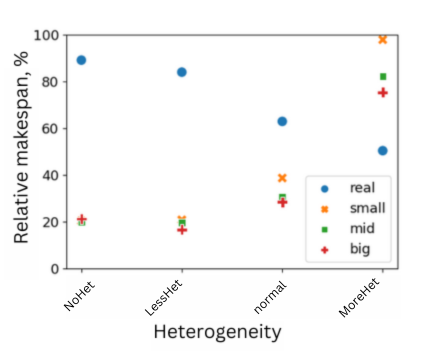}
  \hfill       \includegraphics[width=0.495\columnwidth]{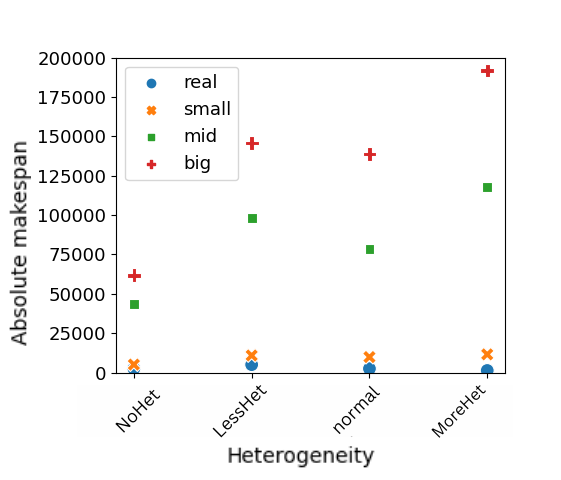}
%  \vspace{-0.3cm}
  \caption[]{Relative (left, baseline: \dagmem) and absolute (right) makespan of \daghetpart for different levels of  heterogeneity. Smaller is better.}
  \Description{Relative (left, baseline: \dagmem) and absolute (right) makespan of \daghetpart for different levels of  heterogeneity. Smaller is better.}
  \label{fig:improvementbyheterogeneity}       \label{fig:improvementbyheterogeneityabsolute}
 % \vspace{-0.3cm}
\end{figure}

Fig.~\ref{fig:improvementbyheterogeneity} (left) illustrates the impact
of the heterogeneity level on the relative makespan.
Somewhat unintuitively, with more hete\-rogeneity, the relative makespans grow,
i.e., \daghetpart wins against the baseline \dagmem by a smaller margin, except for real-world workflows.
The absolute makespan values of \daghetpart have a similar tendency to grow with more heterogeneity (Fig.~\ref{fig:improvementbyheterogeneityabsolute}, right).
Note that this is not because \daghetpart works better in a less heterogeneous environment.
The reason lies in the cluster configuration: clusters with less heterogeneity still have to be able to fit even the most memory-demanding task.
Hence, they still have the {\em luxury} processors $C2$, as well as others that are modified to be more similar to them.
In case of no heterogeneity, all processors are $C2$.
All the parts that were earlier processed by slow processors are now being processed by~$C2$ processors or similar ones.
In case of more heterogeneity, on the other hand, the rest of the cluster is weaker in comparison.
The parts that are assigned to weaker processors take longer to execute.

On top of that, the baseline receives a benefit in our more heterogeneous clusters when choosing the processors,
since it starts with the processor with the largest memory.
In our cluster with more heterogeneity, this processor has not only more memory in relation to the others (compared to the
normal scenario), but also the highest speed. % , increased even more for this scenario.
% This makes mapping tasks on it even more advantageous.
The strategy of \dagmem to map big parts on these first processors pays off more
than in the default cluster, so that it becomes harder for \daghetpart to improve this baseline.
Still, regardless of the level of heterogeneity and the type of workflow,
\daghetpart improves over the baseline in all cases.
%    on the baseline by $3.7\%$ on small workflows, by $17.8\%$ and by $24.8\%$ on big workflows.

%Svetlana: I commented this out, because it dopesn't fit the narrative anymore
%In the NoHet case, the relative improvement of all workflows fits in the general trend: on workflows whose size allows \daghetpart to properly
%utilize the parallelism, the improvement of \daghetpart over \dagmem is around $5\times$ (the makespans are $19$-$21\%$ of those
%produced by \dagmem).

As already mentioned, real-world workflows show a different trend from the generated ones.
The relative makespan further improves with an increased level of heterogeneity.
%    The makespan improvement by \daghetpart grows with growing heterogeneity.
%    On a cluster with less heterogeneity, the improvement is $16.2\%$, while on the cluster with more heterogeneity
%    it is already $49.66\%$.
Here, the small size of the workflows plays a positive role: the workflows are mapped exclusively on the fastest
nodes, so when these nodes are stronger, then \daghetpart, a heuristic that is able to better utilize parallelism, wins.
If the size of the workflow requires the use of weaker nodes, then the benefit provided by very fast nodes becomes less pronounced for \daghetpart.

  \subsubsection{Impact of computational demands -- Workflows with four times bigger $w_u$}
  To explore the impact of higher computational demands, we evaluate workflows that deviate only in the workloads of their tasks $w_u$.
  Thus, for the generated workflows, we assign four times larger workloads; for the real-world ones, we multiply the historical values by four.
  Relative makespans in both cases are virtually identical.
  For example, for real-world workflows, \daghetpart produces makespans that are $62.8\%$ of those produced by \dagmem ($1.59\times$ better).
  On real-world workflows with increased computational demands, \daghetpart produces an average makespan of $61.73\%$ ($1.62\times$ better).
  Small, \new{middle}-sized and big workflows yield average makespans of $36.4\%$, $30.9\%$, and $30.3\%$ on high-demand workflows, respectively, while normal-demand workflows yield $38.6\%$, $30.63\%$, and $28.4\%$.
  This means that a symmetrical increase in computational demands of all tasks has little impact on \daghetpart's quality.
%  \todo{Maybe move most of this to appendix}
%\skug{maybe the rest of the paragraph is not necessary}
%However, absolute makespans increase: for generated workflow approximately $3.5$-fold (\eg from $138885$ to $479092$ for big workflows),
%for real-world workflows approximately $3$-fold (from $2493$ to $6988$).
%Seemingly, a part of the $4$-fold increase in computational demands is optimized out by both \dagmem and \daghetpart,
%therefore absolute makespans of \daghetpart that are less than $4$ times bigger are still in the same relation to makespans
%produced by \dagmem.

{
  \subsubsection{Behavior of workflow types when scaling workflow size}
  Previously, we presented geometric means of makespans among workflows in a certain size category (\eg big or small).
% \hmey{Why geomeans of absolute numbers? Because they are of different scales?
% Are those means then still expressive enough?}
  Yet, each category consists of workflows of different kinds, depending on the model workflow that was used for their generation.
  For example, the ``small'' category contains the ``Epigenomics'' workflow with  $2~000$ tasks, while ``big'' contains the scaled one with ${30~000}$ tasks.
  Hence, in this section we present unaggregated results from \daghetpart on each workflow family, varied by workflow size.

  Fig.~\ref{fig:bytyperel} shows, for each workflow type, the relative makespans in relation to the size.
  The solution quality of \daghetpart varies here across different families of workflows.
  There are workflows that are consistently easy for \daghetpart to map, \eg Seismology, BWA and BLAST.
  These are workflows with the highest degrees and fanout.
  There are workflow families where \daghetpart performs better with increasing size of the workflow, like Genome and Soykb
  (note that for families like Soykb or Montage, not all workflow sizes are present).
  Especially on small Soykb workflows, the improvement achieved by \daghetpart is smaller than usual (less than $20\%$).
  Soykb starts with a chain of tasks and ends with a fork-join segment.
  With growing size, however, there is more parallelism to be utilized, and hence \daghetpart is able to improve more over \dagmem.

  \begin{figure}
    %\vspace{-.4cm}
      \centering
      \includegraphics[width=0.95\columnwidth]{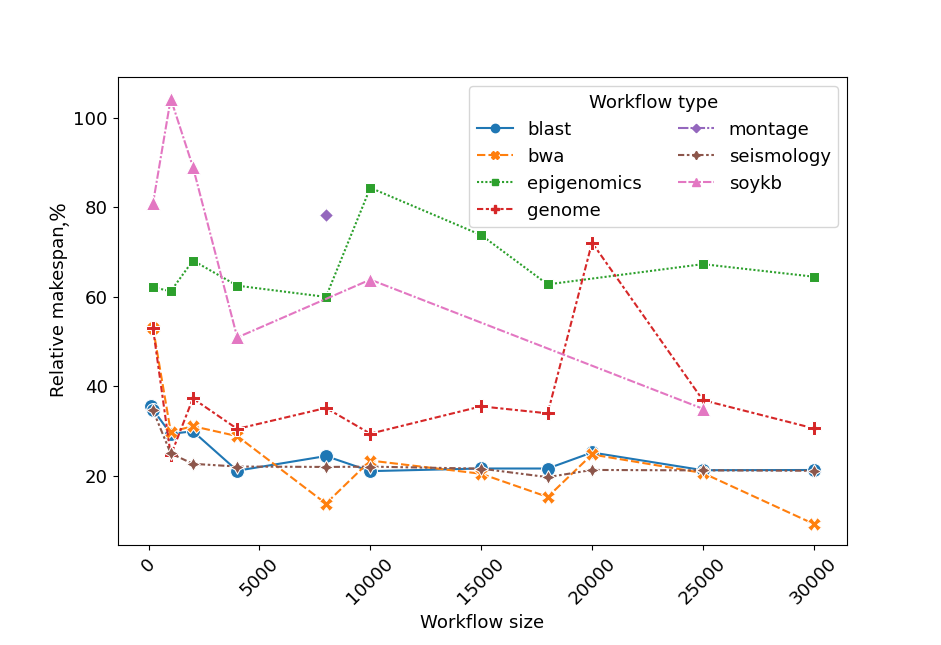}
   % \vspace{-.4cm}
    \caption[]{Makespan of \daghetpart relative to \dagmem for different workflow families.
    {Dotted lines are meant to improve readability.}}
    \Description{Makespan of \daghetpart relative to \dagmem for different workflow families.
        Dotted lines are meant to improve readability.}
    \label{fig:bytyperel}
  \end{figure}

\new{
  Absolute makespans per workflow family are presented in Fig.~\ref{fig:bytypeabs}.
  For all families except two, the makespans seem to grow in a roughly linear manner.
  For the two difficult families soykb and epigenomics, the makespans seem to grow faster.
  Yet, considering that the relative makespans for these families fall slightly with the increasing
  workflow size, the superlinear growth of absolute makespans is apparently a feature of the workflow, not of \daghetpart's quality.

  \begin{figure}[htb]
   % \vspace{-.1cm}
      \centering
      \includegraphics[width=0.95\columnwidth]{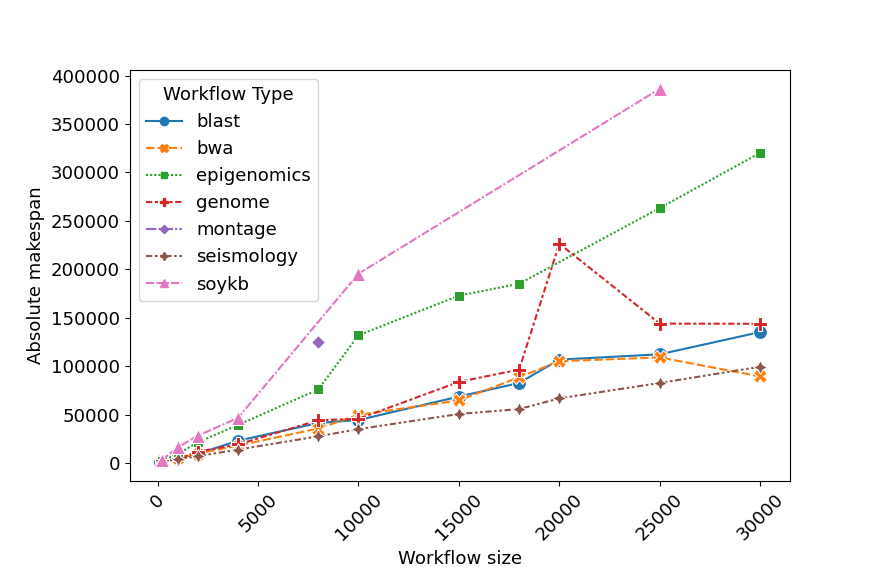}
   % \vspace{-.3cm}
    \caption[]{Absolute  makespan of \daghetpart for different families of workflows.
    Dotted lines are meant to improve readability.}
    \Description{Absolute  makespan of \daghetpart for different families of workflows.
        Dotted lines are meant to improve readability.}
    \label{fig:bytypeabs}
  %  \vspace{-.1cm}
  \end{figure}
    }

\subsubsection{Impact of communication-to-computation ratio (CCR)}

Fig.~\ref{fig:improvementbybeta} shows the effect of different cluster bandwidths on the relative performance of
\daghetpart compared to the baseline.
The general trend shows that higher bandwidths allow \daghetpart to better exploit
heterogeneity and produce better makespans.
In particular, on small workflows, the relative makespan decreases by $13$ percentage points from the smallest bandwidth to the largest one.
For the other two groups, the effect is smaller: for  big workflows, the difference between the worst and the best relative makespan
is only $5.3$ percentage points.
Real-world workflows prove resistant to changes in bandwidth, because they occupy fewer nodes and the
communication costs matter less.

\begin{figure}[tb]
  \centering
  \includegraphics[scale=0.48]{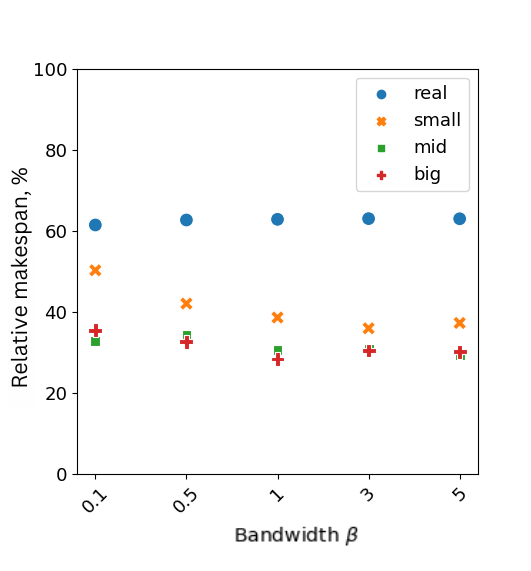}
  %\vspace{-.6cm}
  \caption[]{Relative makespan of \daghetpart compared to \dagmem on default cluster, in percentage, as a function of  bandwidth. Smaller is better.}
  \Description{Relative makespan of \daghetpart compared to DagHetMem on default cluster, in percentage, as a function of  bandwidth. Smaller is better.}
  \label{fig:improvementbybeta}
 % \vspace{-.3cm}
\end{figure}

Blocks are based on the partition provided by the \algo{dagP} algorithm, which optimizes for edge cut.
Even though we merge and repartition the blocks further, this effect of low communication costs seems to persist.
This explains the relatively small reaction of \daghetpart makespan on varying bandwidths.

\new{
However, different families of workflows react differently to changes in the bandwidth.
On the two most fanned-out families (BWA and Blast), \daghetpart provides $3.14$ times smaller makespans
on small workflows in a cluster with the biggest bandwidth in comparison to the smallest bandwidth.
On the two least fanned-out families (Soykb and Epigenomics), the makespans in the same conditions are
only $1.27$ times smaller than those on the smallest bandwidths.
For the middle-sized workflows, the result for the most fanned-out families $3.31\times$ better, while the least fanned-out
ones yield $1.22\times$ smaller makespans on the biggest bandwidths.
On big workflows with the biggest bandwidth, the two most fanned-out families lead to makespans that are $3.3\times$ better compared to the results for the smallest bandwidth.
The two least fanned-out ones, in turn, only lead to makespans that are $1.4\times$ better on average
in this comparison.
}

\subsubsection{Running times}
Both algorithms need on real-world workflows less than one second on average. That is why it
is of no concern that \dagmem is much faster in this category.
Also, the small workflows can be handled in a few seconds by both heuristics,
and \daghetpart is only $63\%$ slower than
\dagmem. On middle-sized workflows, both algorithms are nearly equally fast and require less than three minutes on
average. \daghetpart is even faster than \dagmem on big workflows;
on average, it can map them in less than $11$~minutes and thus scales even to big workflows in very reasonable time.

{The running time of a mapping/scheduling algorithm has to be seen in the context of the running time improvement of the resulting schedule.
For real-world workflows, the geometric mean of the makespan improvement corresponds to $5.3$ minutes, while the average absolute
scheduling running time is around $0.5$ seconds.
The makespan improvement can correspond to up to $97.8$ hours for large workflows (methylseq).
For synthetic workflows (small/middle/big), scheduling time is on average $2.68$/$166$/$647$ \emph{seconds};
the resulting makespan improvements correspond to $10.9$/$41.8$/$84.3$ \emph{hours}.
On all categories of workflows, this is a significant gain and a good reason to use an advanced mapping/scheduling method.
}

In fact, the running time of \dagmem is dominated by the effort to compute the optimal memory traversal over the entire workflow.
With increasing workflow size, this procedure becomes more and more time-consuming.
\daghetpart computes optimal memory traversals only on the blocks that are smaller in size.
Therefore, for bigger workflows, its performance is better than that of \dagmem.
Table~\ref{tab:runtimes} shows relative and absolute running times of \daghetpart, where the relative ones are
normalized by those of \dagmem.
Fig.~\ref{fig:runtimessize} shows the relative running time of \daghetpart for each workflow,
again relative to \dagmem.
Finally, Fig.~\ref{fig:runtimessize2} reports {absolute running times, confirming the trend.
The logarithmic y-axis lowers the perceived variance for the big workflows, but allows the reader to grasp the variance of the
smallest workflows.}

\begin{table}[htb]
\begin{center}
\begin{tabular}{rcc}
\toprule
Workflow set  &  avg.\ relative runtime & avg.\ absolute runtime (sec)  \cr
\midrule
real-world               & 406   & 0.5    \cr
small & 1.63 & 2.83\cr
middle & 1.02 & 166.39\cr
big & 0.85 & 647.13 \cr
\bottomrule\end{tabular}\end{center}
\caption[]{Relative (compared to \dagmem) and absolute running times of \daghetpart.}
\label{tab:runtimes}
\end{table}

\begin{figure}[htb]
\centering
\includegraphics[scale=0.2]{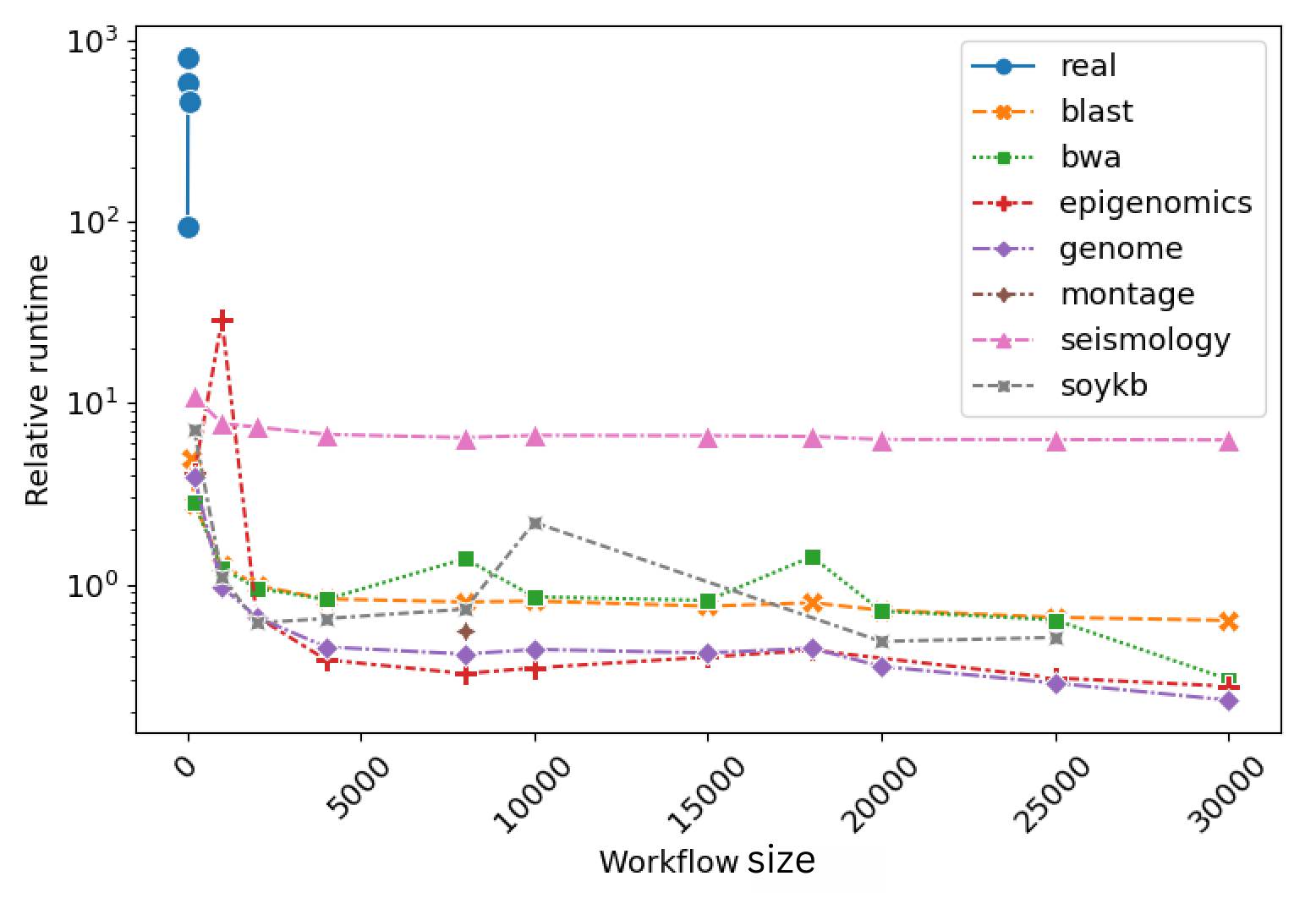}
%\vspace{-.4cm}
\caption[]{Rel. running time of \daghetpart to \dagmem.
{Dotted lines are meant to improve readability.}}
\Description{Rel. running time of DagHetPart to DagHetMem.
Dotted lines are meant to improve readability.}
\label{fig:runtimessize}
%\vspace{-.3cm}
\end{figure}

\begin{figure}[htb]
\centering
\includegraphics[scale=0.21]{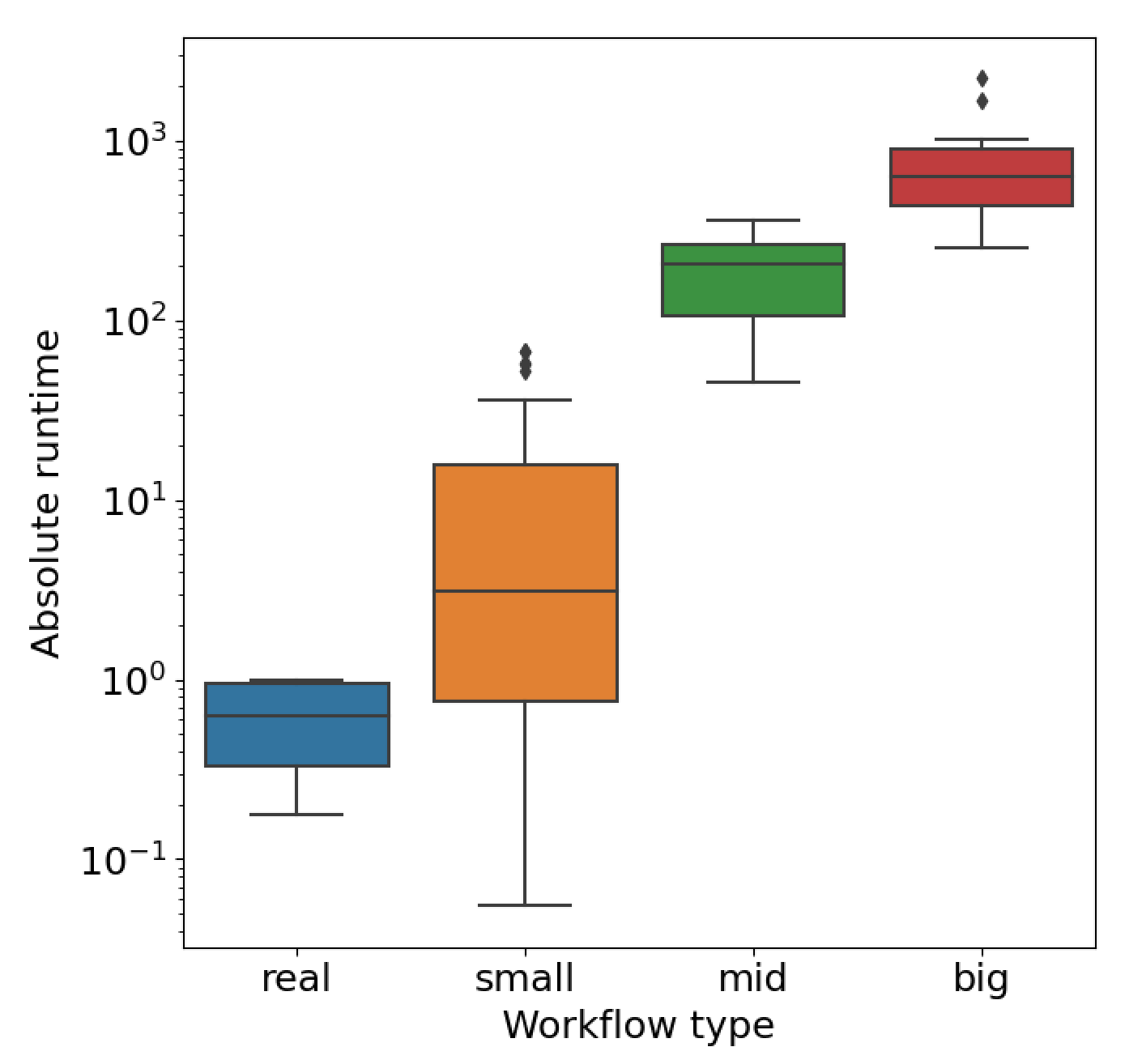}
\caption[]{Absolute running time of \daghetpart (by workflow types). {Note that the y-axis has a logarithmic scale.} }
\Description{Absolute running time of DagHetPart (by workflow types). Note that the y-axis has a logarithmic scale.}
\label{fig:runtimessize2}
%\vspace{-.3cm}
\end{figure}

\subsubsection{Summary}
Overall, \daghetpart is improving makespans upon the baseline for all workflow sizes
and on all cluster configurations, even though the gain is affected by
the cluster configuration.
In particular, with a large cluster, there is a significant gain, which grows
to an improvement of $4\times$ to $5\times$ compared to the baseline (except
for real-world workflows that are smaller in size).
{While the relative makespan increases when the heterogeneity is increased,
  we always observe a consistent improvement over the baseline, even in a
  fully homogeneous setting. Also, increasing the computational demand leads
  to similar relative values of makespan.}
Finally, increasing the communication bandwidth in the cluster has a positive impact
on the improvement of the makespan of \daghetpart, which better exploits the workflow parallelism
compared to \dagmem.
\new{
\daghetpart employs more processors and requires more communication.
When the bandwidth grows, it benefits more from a high throughput of the execution environment}.
}

\section{Conclusions} 
\label{sec:conc}
We have tackled the fundamental but very complex problem of mapping  large-scale
memory-constrained workflows, modeled as a directed acyclic graph, onto a
heterogeneous platform where each processor may have a different memory capacity
and a different processing speed.
In this setting, the objective is to minimize the total execution time (makespan).
% \iec to strive for the best possible performance when executing the workflow.
One of the main challenges % faced on the path to performance
is that % workflows are memory-intensive and
some processors may not have enough memory to execute the workflow part
assigned to them. While several mapping and scheduling algorithms have been proposed
in various models, the problem of makespan minimization while accounting
for memory constraints on DAGs has never been addressed, \new{to the best of our knowledge}.

To palliate the lack of heuristics accounting for memory constraints, we first proposed
a baseline heuristic, \dagmem, building upon existing work,
  that produces a valid mapping. Next, we  designed a sophisticated
new four-step heuristic, \daghetpart, which gradually improves a mapping solution that respects
all constraints. Extensive simulation experiments
on diverse workflows demonstrate that it pays off significantly
to account for the platform heterogeneity and to exploit the parallelism available.
On average, the makespan achieved by \daghetpart is a factor of $2.44$ smaller
than the baseline's, and it may become up to $5$ times smaller for big workflows
and large clusters.
This significant improvement as well as its good scalability to big workflows are
strong indications that the heuristic should work well in practical
scenarios.
%  will be devoted to performing real experiments to confirm this trend.
As future work, we plan to perform more real-world experiments
to confirm this trend, and
to  add one more level of heterogeneity by considering
%consider
% communication links with
different communication bandwidths.

\bibliographystyle{ACM-Reference-Format}
\bibliography{references}

%\clearpage
%\appendix

\end{document}

%% file: macros.tex
\newcommand{\NP}{$\mathcal{NP}$}

\newcommand{\iec}{\textit{i.\,e.},\xspace}

\newcommand{\egc}{\textit{e.\,g.},\xspace}
\newcommand{\eg}{\textit{e.\,g.}\xspace}

\newcommand{\etal}{\textit{et al.}\xspace}

\newcommand{\wrt}{w.\,r.\,t.\xspace}